# Ferroelectric Transition in Compressively Strained SrTiO$_3$ Thin Films

Amit Verma[1,2*], Santosh Raghavan[3], Susanne Stemmer[3] and Debdeep Jena[1,2,4]

[1]Department of Electrical Engineering, University of Notre Dame,
Notre Dame, Indiana 46556, U.S.A.
[2]School of Electrical and Computer Engineering, Cornell University,
Ithaca, New York 14853, U.S.A.
[3]Materials Department, University of California, Santa Barbara, California 93106, U.S.A.
[4]Department of Materials Science and Engineering, Cornell University,
Ithaca, New York 14853, U.S.A.
*E-mail: averma@cornell.edu

Abstract: We report the temperature dependent capacitance-voltage characteristics of Pt/SrTiO$_3$ Schottky diodes fabricated using compressively strained SrTiO$_3$ thin films grown on (LaAlO$_3$)$_{0.3}$(Sr$_2$AlTaO$_6$)$_{0.7}$ (LSAT) substrates. The measurements reveal a divergence of the out of plane dielectric constant of SrTiO$_3$ peaked at ~140K, implying a ferroelectric transition. A Curie-Weiss law fit to the zero-bias dielectric constant suggests a Curie temperature of ~56 K. This observation provides experimental confirmation of the theoretical prediction of out of plane ferroelectricity in compressively strained SrTiO$_3$ thin films grown on LSAT substrate. We also discuss the roles of the field-dependent dielectric constant and the interfacial layer in SrTiO$_3$ on the extraction of the Curie temperature.

Ferroelectric crystals exhibit a spontaneous non-zero electric dipole moment below a certain transition temperature (Curie temperature $T_C$) due to inversion symmetry breaking [1]. Above this transition temperature, the crystal is in a paraelectric state with zero spontaneous electric dipole moment. This temperature-driven structural phase transition is accompanied by a divergence in the dielectric constant. As the temperature is decreased, the dielectric constant of a ferroelectric crystal increases in accordance with the Curie-Weiss law and peaks at the transition temperature [1, 2]. The dielectric constant decreases from this peak value on further reducing the temperature.

SrTiO$_3$ is a transition metal oxide that crystallizes in a cubic perovskite crystal structure. Many commercially used ferroelectric materials such as BaTiO$_3$, PbTiO$_3$, Pb(Zr$_x$Ti$_{1-x}$)O$_3$ (PZT) also crystallize in this crystal structure [3]. Similar to these well-known ferroelectrics, the dielectric constant of bulk SrTiO$_3$ increases as the temperature is lowered, following the Curie-Weiss law [4]. However, in contrast to traditional ferroelectrics where the dielectric constant peaks and then decreases with temperature, in SrTiO$_3$ the dielectric constant saturates below ~4K [4,5]. Because of this unique behavior, SrTiO$_3$ is often termed as an incipient ferroelectric. This very low temperature dielectric behavior arises because of the quantum fluctuations in SrTiO$_3$ and a preceding antiferrodistortive structural phase transition [5,6]. Theoretical calculations based on the Landau-Ginzburg-Devonshire theory have suggested that the ferroelectric state can be stabilized in SrTiO$_3$ by applying a biaxial tensile or compressive strain [7]. According to these predictions, a tensile strained (001) SrTiO$_3$ crystal should exhibit ferroelectricity in the in-plane direction and a compressively strained (001) SrTiO$_3$ crystal in the out-of-plane direction [7,8]. Such biaxial strains can readily be achieved by heteroepitaxial growth of SrTiO$_3$ on lattice-mismatched substrates [8,9]. In agreement with the theoretical predictions, near room temperature ferroelectricity in the in-plane direction was discovered in tensile strained SrTiO$_3$ thin films grown on DyScO$_3$ substrates [8]. In this earlier study, the capacitance of planar interdigitated capacitors was measured as a function of temperature to extract the temperature dependence of the in-plane dielectric constant [8]. This earlier theoretical work also predicted a divergence in the out of plane dielectric constant in *compressively* strained SrTiO$_3$ thin films grown on (LaAlO$_3$)$_{0.3}$(Sr$_2$AlTaO$_6$)$_{0.7}$ (LSAT) substrates with a $T_C$ of ~50-200K [8]. Till date, there is no experimental evidence of such divergence for compressively strained SrTiO$_3$.

In this work, we discover a strong signature of out-of-plane ferroelectricity in recently reported Pt/SrTiO$_3$ Schottky diodes fabricated on compressively strained SrTiO$_3$ thin films on LSAT substrates [10]. The depletion capacitance of a Schottky diode is a function of the out-of-plane dielectric constant of the material [11]. To capture the T-dependence of the out-of-plane dielectric constant of SrTiO$_3$, we have performed temperature dependent capacitance-voltage measurements on Pt/SrTiO$_3$ Schottky diodes. We find that the zero-bias Schottky depletion capacitance peaks at ~140K. The corresponding divergence in the out of plane dielectric constant implies a ferroelectric transition in the compressively strained SrTiO$_3$ film. The presence of this ferroelectric transition and the extracted Curie temperature are in agreement with earlier theoretical predictions [7,8].

For testing the out-of-plane ferroelectricity in compressively strained SrTiO$_3$, we grew 160nm thick SrTiO$_3$ thin films on a (001) LSAT substrate using hybrid molecular beam epitaxy (MBE) [10]. In this growth technique, the organometallic precursor titanium tetra isopropoxide (TTIP) is used to provide Ti and O, while Sr is provided using an effusion cell [12,13]. For the growth, Sr and TTIP beam equivalent pressures of 6.5 x $10^{-8}$ Torr and 2.3x$10^{-6}$ Torr were used respectively. The background pressure during the growth was ~2x$10^{-8}$ Torr. The growth was performed at a substrate temperature of 900C for one hour at a growth rate ~160 nm/hr. In the MBE system, additional O can be provided during the growth using an oxygen plasma source to make the films insulating. However, to realize Schottky diode devices, the SrTiO$_3$ films used in this study were grown without oxygen plasma. The resultant slightly oxygen deficient conditions dope the SrTiO$_3$ thin films n-type with an electron concentration of $N_D$ ~$10^{19}$ cm$^{-3}$. Since the conditions remained the same throughout the growth, uniform carrier concentration is expected in the film. Using Hall-effect measurements performed in a Van der Pauw geometry, a sheet electron concentration of ~1.28 x $10^{14}$ cm$^{-2}$ was obtained [10]. This sheet concentration value is lower than the expected value of ~1.6 x $10^{14}$ cm$^{-2}$ because of the surface depletion in the SrTiO$_3$ thin film [14, 15].

Both SrTiO$_3$ (STO) and LSAT substrates have a cubic perovskite crystal structure with lattice constants of $a_{STO}$ = 3.905 Å and $a_{LSAT}$ = 3.868 Å, respectively [16]. A recent growth study using the same hybrid MBE technique has found the critical thickness of SrTiO$_3$ thin films grown on LSAT substrate to be ~180 nm [16]. Below this critical thickness, SrTiO$_3$ thin films are expected to grow coherently strained to the LSAT substrate with a compressive strain of ($a_{LSAT}$ - $a_{STO}$)/$a_{STO}$ = -0.95%. The oxygen vacancy concentration corresponding to ~$10^{19}$ cm$^{-3}$ electron concentration in our sample is quite low to cause any significant strain effects and all of the strain in the thin film is expected to arise from the lattice mismatch between the SrTiO$_3$ thin film and the LSAT substrate [17]. To confirm growth of stoichiometric SrTiO$_3$ coherently strained to the LSAT substrate, we performed both (002) on-axis 2θ-ω scan and (013) off-axis reciprocal space mapping (RSM) in a high-resolution Philips Panalytical X'Pert Pro thin-film diffractometer using Cu K$_α$ radiation. The results of the measurements are shown in Figs. 1(a) and 1(b), respectively. The SrTiO$_3$ out of plane lattice parameter, as measured from the 2θ-ω scan (Fig. 1(a)) is $a_⊥$ = 3.932 ± 0.001 Å, as expected from a completely coherent stoichiometric film [16]. The RSM of the sample (Fig. 1(b)) shows that both the SrTiO$_3$ thin film and the LSAT substrate have the same in-plane lattice parameter, further confirming the pseudomorphic film growth and the desired compressive strain in the SrTiO$_3$ layer.

To measure the out-of-plane dielectric constant of the compressively strained SrTiO$_3$ thin film, we fabricated circular Schottky diodes. Optical photolithography was used to pattern the film. Al/Ni/Au (40/40/100 nm) ohmic contacts and Pt/Au (40/100nm) Schottky contacts were deposited using e-beam evaporation. To reduce gate leakage and improve rectification, an Oxygen plasma treatment of the SrTiO$_3$ surface was performed in a reactive ion etching system prior to the Pt/Au Schottky metal deposition. More details on the device fabrication process have been reported elsewhere [10].

For measuring the Schottky diode current-voltage (I-V) and capacitance-voltage (C-V) characteristics, a Keithley 4200 semiconductor characterization system was used along with a Cascade probe station for room temperature measurements and a Lakeshore probe station for temperature dependent measurements. Rectifying I-V characteristics of a Pt/SrTiO$_3$ circular Schottky diode of 10μm radius measured at room temperature are shown in Fig. 2. From the forward bias characteristics, the barrier height for Pt was found to be ~0.86 eV with an ideality factor of $n$ ~1.63. Temperature dependent C-V characteristics (frequency 100 kHz, signal amplitude 30 mV) of the Schottky diode are shown in Fig. 3(a) (80K-140K) and Fig. 3(b) (140K-400K). Near zero bias, where the loss is low and the capacitance extraction using a parallel R-C equivalent circuit is valid, the capacitance increases from 80K to 140K (Fig. 3(a)). On further increasing the temperature beyond 140K, however, the measured capacitance decreases (Fig. 3(b)). All measured devices exhibited similar capacitance behavior as a function of temperature. To depict this capacitance variation more clearly, the zero bias capacitance is plotted as a function of the sample temperature in Fig. 3(c). The capacitance peaks at ~140K, increasing more than 60% compared to the value at 400K.

The depletion capacitance of a Schottky diode is given as $C_d(V) = \sqrt{q\varepsilon_0\varepsilon_r N_D / 2V}$, where $q$ is the electron charge, $\varepsilon_0$ is the vacuum permittivity, $\varepsilon_r$ is the out-of plane dielectric constant of SrTiO$_3$, $N_D = 10^{19}$ cm$^{-3}$ is the doping density, and $V$ is the total voltage drop across the Schottky depletion region [11]. In traditional semiconductors, carriers can freeze out at low temperatures but because of the large dielectric constant of SrTiO$_3$, impurity doped (La or Nb) or Oxygen vacancy doped carriers in SrTiO$_3$ do not freeze out even at liquid helium temperatures [18-20]. Therefore, any temperature dependence of measured Schottky capacitance in our devices should arise from temperature dependence of SrTiO$_3$ dielectric constant. Using the one to one mapping between the measured depletion capacitance and the out-of plane dielectric constant, we can extract the temperature dependence of the dielectric constant. As $\varepsilon_r \propto C_d^2$, the capacitance peak at 140K would imply a peak in SrTiO$_3$ out-of plane dielectric constant at the same temperature. For the zero bias case $V$=0.86 V due to the built-in bias due to Pt. The out-of-plane dielectric constant extracted from the zero bias measured capacitance value is shown in Fig. 4(a). Compared to bulk unstrained SrTiO$_3$ where the dielectric constant saturates at low temperatures [4,5], the divergence in out-of plane dielectric constant observed in the compressively strained SrTiO$_3$ thin films suggests a ferroelectric transition. A Curie-Weiss law fit to $1000/\varepsilon_r$ is shown in Fig. 4(b). From this fit, the extracted Curie temperature, $T_C$, for the SrTiO$_3$ ferroelectric transition is ~56K. This $T_C$ value lies towards the lower limit of the range of theoretically predicted ferroelectric transition temperatures in the SrTiO$_3$/LSAT system (~50-200K) [8]. Our temperature dependent C-V results therefore confirm the theoretical predictions of out-of plane ferroelectricity in compressively strained SrTiO$_3$ thin films grown on LSAT substrate.

Compared to the peak dielectric constants observed for in-plane ferroelectricity in SrTiO$_3$ [8], the dielectric constants observed in this study are one order lower. A possible reason for lower dielectric constants measured is the presence of non-zero electric fields in the Schottky depletion region at zero bias; these fields were absent in the earlier studies that did not use Schottky diodes. Even modest dc electric fields of the order of ~10 kV/cm can drastically reduce the dielectric constant of a ferroelectric material [8] by Coulomb-clamping the motion of the ionic crystal responsible for high dielectric constants in ferroelectrics. The peak electric fields in our Schottky diode devices are certainly larger than 10 kV/cm. In addition to reducing the measured dielectric constant, non-zero electric fields can also cause a shift in dielectric constant vs temperature curves, moving the peak divergence to higher temperatures and the apparent $T_C$ to lower temperatures [21]. We can estimate the order of magnitude of the shift in $T_C$ due to the non-zero electric field using the Landau-Ginzburg-Devonshire theory of ferroelectrics. For this estimate, assuming a uniform electric field $E$ in SrTiO$_3$, the expansion of free energy $F$ in terms of the polarization order parameter $P$ can be written as [21],

$$F = -EP + g_0 + \frac{\gamma(T - T_{c,0})}{2}P^2 + \frac{g_4}{4}P^4 \quad , \qquad (1)$$

where $g_0, \gamma, g_4$ are constants specific to the ferroelectric and $T_{c,0}$ is the actual ferroelectric transition temperature in the absence of an electric field. The value of the equilibrium polarization can be found by minimizing $F$ with respect to $P$, $\partial F/\partial P = 0$ leading to a relation between $E$ and $P$, $E = \gamma(T - T_{c,0})P + g_4 P^3$. Because of the large dielectric constant values in a ferroelectric, the dielectric constant can be approximated by the dielectric susceptibility $\chi$ as

$$\varepsilon_r = 1 + \chi \approx \chi = \frac{1}{\varepsilon_0}\frac{\partial P}{\partial E} = \frac{1/\gamma\varepsilon_0}{T - (T_{c,0} - 3g_4 P^2/\gamma)} \quad . \quad (2)$$

The apparent reduction in $T_C$ due to the non-zero electric field is $3g_4 P^2/\gamma$. Comparing Eq. (2) to the fit $\varepsilon_r = 36200/(T-56)$ shown in Fig. 4(b), we estimate the value of $\gamma$ to be $\gamma \sim 3.12 \times 10^6$ m$^2$N/C$^2$. The linear fit to experimental data (Fig.4(b)) is good for T ~ 250K-400K. In this temperature range, $P$ can be approximated as $P \sim E/\gamma(T - T_{c,0})$. To get an estimate of $3g_4 P^2/\gamma$, we can take $g_4 \sim 6.8 \times 10^9$ m$^6$N/C$^4$, equal to the bulk unstrained SrTiO$_3$ value [22], and $(T - T_{c,0}) \sim 100$K. For an average electric field of ~100 kV/cm in SrTiO$_3$, the apparent reduction in $T_C$ would then be $3g_4 P^2/\gamma \sim 7$K.

In our experiment, the extracted $T_C$ value of ~56K directly depends on the measured capacitance. The presence of an interfacial low dielectric constant layer (also called dielectric dead layer) at the Schottky metal/SrTiO$_3$ interface has been frequently reported in the literature [23-28]. The capacitance $C_i$ of such an interfacial layer acts in series with the Schottky depletion capacitance $C_d$ (Inset: Fig.4(a)), thus reducing the measured capacitance $C_m = C_d C_i/(C_d + C_i)$ and hence the extracted dielectric constant [23-28]. Since $\varepsilon_r \propto 1/(T - T_C)$, this reduction in extracted dielectric constant would show up as an apparent reduction in $T_C$, similar to the case of non-zero electric field in SrTiO$_3$. To find this $T_C$ shift, we need an estimate of $C_i$.

$C_i$ and $C_d$ are related to the ideality factor $n$ of the Schottky diode as $C_i/C_d = 1/(n-1)$ [28]. Since, $C_m = C_d C_i/(C_d + C_i)$, we can also write this capacitance ratio in terms of the measured capacitance $C_m$ and interfacial capacitance $C_i$ as $C_i/C_d = C_i/C_m - 1$. Combining these two expressions, we can directly obtain $C_i$ from the ideality factor $n$ (~1.63) and the measured capacitance $C_m$ as $C_i = nC_m/(n-1)$. At zero applied bias, the measured Schottky diode capacitance at room temperature is $C_m \sim 3.56$ µF/cm$^2$, therefore $C_i = nC_m/(n-1) \sim 9.21$ µF/cm$^2$. Assuming that the interfacial capacitance is independent of temperature, the effect of interfacial capacitance can be de-embedded from the measured capacitance to calculate the actual Schottky depletion capacitance [$C_d = C_i C_m/(C_i - C_m)$] at different temperatures. The out-of-plane dielectric constant extracted from this de-embedded Schottky capacitance is shown in Fig.4 (a) along with the Curie-Weiss law fit [Fig. 4(b)]. After removing the effect of the interfacial capacitance, the extracted Curie temperature, $T_C$, for the SrTiO$_3$ ferroelectric transition is ~141 K. Combining this apparent reduction in $T_C$ due to the interfacial capacitance with the reduction in $T_C$ due to the non-zero electric fields in Schottky diodes, the actual Curie temperature for compressively strained $SrTiO_3$ is expected to be ~148 K, in the upper half of the theoretically predicted range of ~50-200 K.

To summarize, in this work using temperature dependent C-V measurements performed on Pt/$SrTiO_3$/LSAT Schottky diodes fabricated using compressively strained $SrTiO_3$ thin films, we have observed a divergence in the out-of plane dielectric constant in $SrTiO_3$. This finding provides an experimental confirmation of theoretical predictions of out-of-plane ferroelectricity in $SrTiO_3$ thin films coherently strained to LSAT substrate. We hope this work will help in designing and understanding of $SrTiO_3$ based devices on this widely used substrate. In addition to the divergence of the dielectric constant, ferroelectricity is also accompanied with the emergence of field switchable non-zero spontaneous polarization below the Curie temperature. This property is difficult to probe in a Schottky diode geometry, but potentially can be demonstrated in future studies by other techniques.

The authors thank Evgeny Mikheev for useful discussions. This work was supported by the Extreme Electron Concentration Devices (EXEDE) MURI program of the Office of Naval Research (ONR) through grant No.N00014-12-1-0976.

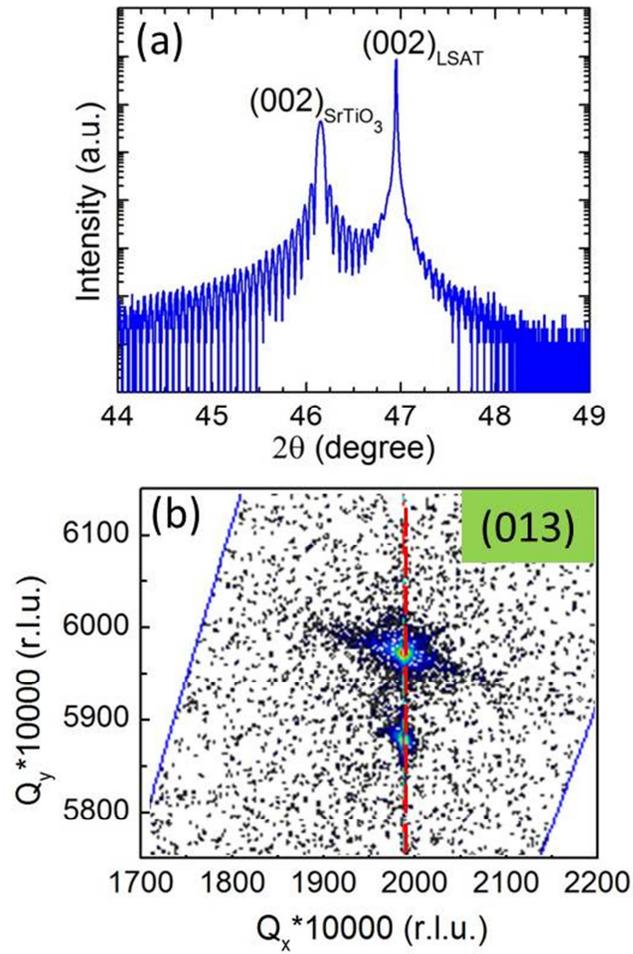

**FIG. 1:** a) (002) on-axis x-ray diffraction 2θ-ω scan of the grown $SrTiO_3$/LSAT sample, (b) (013) off-axis reciprocal space map of the grown sample showing coherently strained pseudomorphic growth of the $SrTiO_3$ thin film on the LSAT substrate.

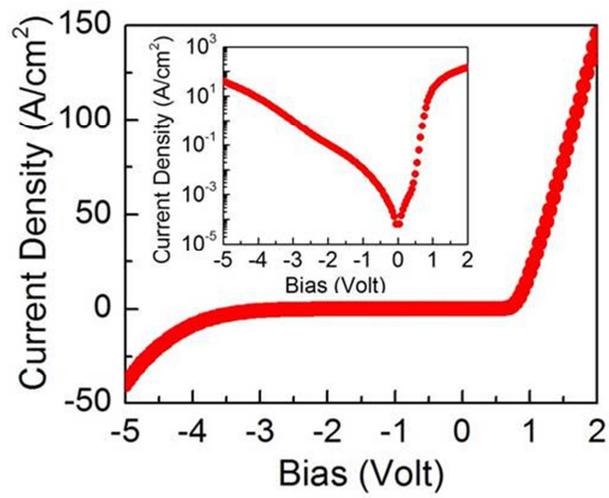

**FIG. 2:** Measured room temperature I-V characteristics of a 10μm radius Pt/SrTiO$_3$/LSAT circular Schottky diode.

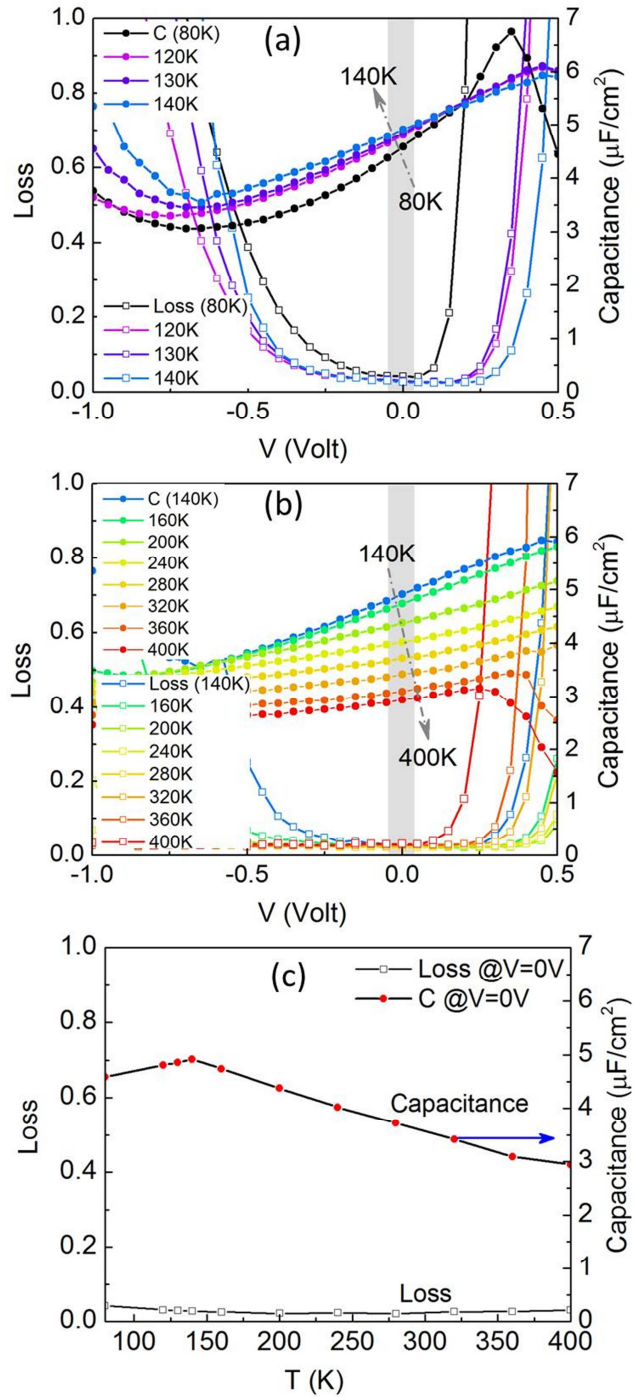

**FIG. 3:** Measured temperature dependent C-V (100kHz, 30mV) characteristics of the 10μm radius Pt/SrTiO$_3$/LSAT circular Schottky diode showing increase in measured depletion capacitance (right axis/filled circles) in the temperature range (a) 80K-140K, followed by a decrease in the capacitance in the temperature range (b) 140K-400K. (c) Variation of zero bias measured capacitance with temperature, capacitance peaks at ~140K. Temperature dependence of loss is also plotted (left-axis/open squares). Loss is quite low around zero bias over the whole 80K-400K temperature range.

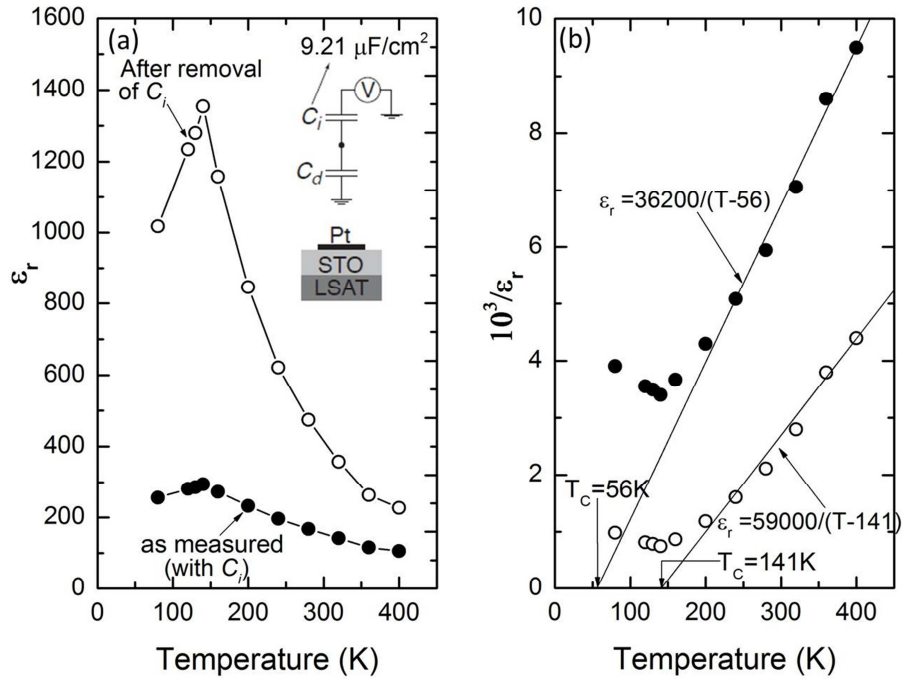

**FIG. 4:** (a) SrTiO$_3$ dielectric constant calculated from the measured zero bias depletion capacitance (solid circles) and after removing the effect of interfacial capacitance $C_i$ (open circles), (Inset: Effective capacitance model of the Pt/SrTiO$_3$ Schottky diode) (b) Curie-Weiss law fit to the inverse of dielectric constant.